\documentclass[showpacs,amsmath,amssymb,floatfix,pre,twocolumn]{revtex4}
\usepackage[latin1]{inputenc}
\usepackage[hiresbb,final]{graphicx}

\newcommand{\dotp}{\mathbin{\boldsymbol{\cdot}}}
\newcommand{\bfnabla}{\boldsymbol\nabla}
\newcommand{\air}{\text{air}}

\newcommand{\bfn}{\mathbf{n}}

\newcommand{\bfv}{\mathbf{v}}
\newcommand{\bfV}{\mathbf{V}}

\begin{document}

\title{Self-consistent theory of capillary-gravity-wave generation by small moving objects}
\author{A. D. Chepelianskii$^\text{(a)}$, M. Schindler$^\text{(b)}$, F.
Chevy$^\text{(c)}$, E. Raphaël$^\text{(b)}$}
\affiliation{\textrm{(a)} Laboratoire de Physique des Solides, Univ. Paris-Sud, CNRS, UMR 8502, F-91405, Orsay, France}
\affiliation{\textrm{(b)} Laboratoire PCT, UMR~\emph{Gulliver} CNRS-ESPCI~7083, 10~Rue Vauquelin, 75231~Paris~Cedex~05, France}
\affiliation{\textrm{(c)} Laboratoire Kastler Brossel, CNRS, UPMC, Ecole Normale Supérieure, 24~rue~Lhomond, 75231~Paris, France}


\begin{abstract}
We investigate theoretically the onset of capillary-gravity waves created by a
small object moving at the water-air interface. It is well established that, for
straight uniform motion, no steady waves appear at velocities below the minimum
phase velocity $c_\text{min} = 23\,{\rm cm/s}$. At higher velocities the
emission of capillary-gravity waves creates an additional drag force. The
behavior of this force near the critical velocity is still poorly understood. A
linear response theory where the object is replaced by an effective pressure
source predicts a singular behavior for the wave drag. However, experimental
data tends to indicate a more continuous transition. In this article, we show
that a proper treatment of the flow equations around the obstacle can regularize
wave emission, even in the linear wave approximation, thereby ensuring a
continuous behavior of the drag force.
\end{abstract}

\maketitle

\section{Introduction}

An object moving uniformly in an incompressible liquid experiences a drag force
that can have several physical origins: viscous drag, hydrodynamic interaction
with close-by boundaries, or the force due to the emission of waves. The waves,
on which we focus in the present paper, appear when the object moves in the
vicinity of a deformable surface such as an air--liquid
interface~\citep{LanLif59}. They carry away momentum from the object which is
sensed as the wave drag of the moving object. The type of waves we expect on an
air-liquid interface are capillary gravity waves~\citep{LanLif59,Acheson}. Their
dispersion relation in an unbounded inviscid liquid of infinite depth is well
known to be $\omega^2 = g k + \gamma k^3 / \rho$. It relates the oscillation
frequency~$\omega$ to the wave number~$k$ and depends on the gravity
constant~$g$, fluid density~$\rho$ and on the surface tension~$\gamma$. The wave
velocity $c(k) = \omega/k$ is readily obtained as $c(k)=(g/k+\gamma
k/\rho)^{1/2}$. The dispersive nature of capillary-gravity waves creates a
complicated wave pattern around a moving object, yielding a finite wave drag. In
naval design the wave drag is an important source of resistance, which
stimulated the development of approximate theoretical
methods~\citep{Lighthill,Lamb,Rayleigh,Kelvin,Milgram,Burghelea1}. These methods
are valid only for objects larger than the capillary length $\kappa^{-1} =
\sqrt{\gamma/(\rho g)}$~\citep{RapGen96}. The case of objects of extension
comparable to~$\kappa^{-1}$ has been overlooked for a long time in the
literature, but has attracted strong interest in the context of insect
locomotion on water surfaces~\citep{Tucker,Denis,Bush,Nachtigall1965}. In
particular, some insect species (for example wiggling beetles) may take
advantage of the generation of capillary-gravity waves for echo-location
purposes~\cite{Denis2,Bendele}. In particular, recent observation of
the behavior of Gyrinidae suggest that they select their swimming speed by
minimizing the sum of wave and viscous drags~\citep{VoiCas09,Alexander,Buhler}.

The first theoretical calculation in this regime predicted a discontinuity of
the wave drag~$R_w$ at a critical velocity given by the minimum of the wave
velocity $c_\text{min} = (4g\gamma/\rho)^{1/4}$ for capillary gravity
waves~\citep{RapGen96}. For water this evaluates to~$c_\text{min}\approx 23
\,\textrm{cm/s}$. An object moving at constant velocity $V < c_\text{min}$ does
not generate steady waves, and the wave resistance vanishes. Emission of steady
waves becomes possible only when $V > c_\text{min}$, leading to the onset of a
finite wave drag. This striking behavior is similar to the well-known Cherenkov
radiation emitted by charged particles~\citep{Cherenkov}, to the onset of wave drag for
supersonic aircrafts~\citep{supersonic} or to the
Zeldovich--Starobinsky effect in general relativity~\citep{Zeldovich}. The minimum in the dispersion
relation, responsible for this behavior, renders the problem challenging. Two
experiments addressed the problem of the behavior of wave resistance at a
liquid--air interface. While the disappearance of wave drag was confirmed for $V
< c_\text{min}$, opposite conclusions were reached concerning the existence of a
discontinuity at the critical velocity $V = c_\text{min}$. In a first experiment
by~\citet{Browaeys} the bending of a narrow fiber in contact with the liquid
surface was used to probe the wave drag. The results evidenced the presence of a
jump at $V = c_\text{min}$. However, during these measurements the contact line
between the fluid and the fiber was free to move, thus creating an uncontrolled
contribution to the measured force. A second experiment was made
by~\citet{BurSte02} in which an ingenious feedback system fixed the immersion
depth of the object. This experiment concluded on a continuous increase of wave
drag around $V = c_\text{min}$.
\begin{figure}%
  \centering
  \includegraphics{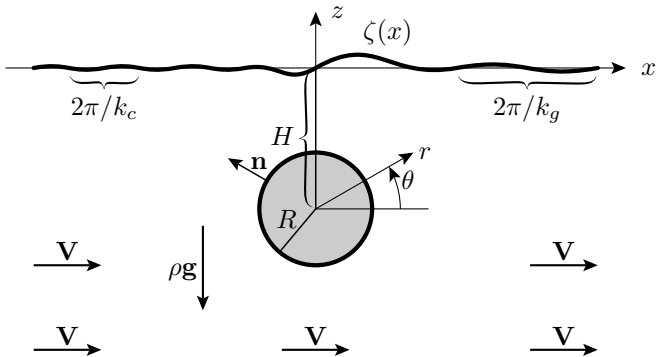}%
  \caption{Geometry of the submerged cylinder and of the free surface above. The
  far-field velocity is indicated by the arrows~$\bfV$.}%
  \label{fig:geometry}%
\end{figure}%

While it was shown recently~\citep{CheCheRap08} that the threshold $V =
c_\text{min}$ exists only for an object moving at constant velocity without
acceleration, a theoretical understanding of the scaling of wave resistance is
still missing. The theoretical descriptions which are available replace the
moving object by an external pressure source applied at the air-liquid
interface. The hydrodynamic problem is then reduced to a linear response theory
in the pressure field which has singularities around $V = c_\text{min}$ if
viscosity is neglected~\citep{RapGen96,Richard,CheRap03}. A self-consistent
determination of the pressure distribution was attempted in
Ref.~\citep{CheRap03}. While this theory succeeded in removing the singularity
at $V = c_\text{min}$, it leads to the somewhat unrealistic prediction that the
applied pressure field vanishes at $V = c_\text{min}$. An alternative approach
by~\citet{Sun} is based on an asymptotic matching technique and works for
velocities much larger than~$c_\text{min}$. In order to develop a theory of wave
drag valid for $V \simeq c_\text{min}$, we note that the linear response theory
is successful at reproducing the wave pattern created in the experiments even
for velocities very close to $c_\text{min}$. This suggests that it is possible
to understand the behavior of~$R_w$ in a theory where linear capillary-gravity
waves are coupled to an accurate hydrodynamic description of the flow around the
moving object. This theory is developed in the present paper.

In the situation we have in mind, the perturbation of the free surface does not
come from an external pressure distribution as it was the case in
Ref.~\citep{RapGen96}. Also, we avoid the difficulties arising from an immersed
perturbing needle of Ref.~\citep{CheRap03}. Instead, we here use a completely
submerged object, which is, for reasons of a dimensionality reduction, a
cylinder. It has radius~$R$ and is at depth~$H$ below the free surface, as is
depicted in Fig.~\ref{fig:geometry}. Notice that the dimension has implications
on the nature of the transition near $c_\text{min}$ which we want to describe
and might limit comparisons with experimental data.

The liquid flows only perpendicularly to the axis of the cylinder, and we
require only $x$~and $z$~coordinates. This system has been analyzed
by~\citet{Lamb} without taking into account the mutual interaction between the
two perturbations, namely the one created by the moving object and the one by
the emitted waves. He limited his discussion to particles larger than the
capillary length~$\kappa^{-1}$ by neglecting the capillary contribution in the
dispersion. We here regard the other case of small objects, where it becomes
necessary to treat the mutual interaction between the two perturbations
correctly.

Generally speaking, the larger the ratio~$H/R$, the better work the two theories
presented below in Secs.~\ref{sec:flow} and~\ref{sec:dipolar}. The numerical
example in Sec.~\ref{sec:num} for $H/R=5/3$, however, shows that the theories
are valid already for cylinders quite close to the surface. This observation is
the reason why we are convinced that the completely submerged cylinder presents
a useful approach to understand the experimental setup in which a partially
immersed object was used.

\section{Flow equations}
\label{sec:flow}

We describe the velocity field in the reference frame of the cylinder. The
far-field velocity is thus a non-zero uniform velocity~$\bfV$. The velocity
field is assumed to be irrotational, given by a velocity potential, where we
immediately isolate the far-field velocity and use only the
potential~$\phi(x,z)$ of the perturbations due to the obstacle and to the waves,
\begin{equation}
  \bfv(x,z) = \bfV + \bfnabla\phi(x,z).
\end{equation}
The incompressibility condition for the fluid reads
\begin{equation}
  \label{incompress}
  \Delta\phi = 0.
\end{equation}

This equation is complemented by boundary conditions at the air-liquid interface
and at the surface of the cylinder. The kinematic boundary condition reads
\begin{equation}
  \label{kinemsphere}
  \bfn\dotp(\bfV + \bfnabla\phi) = 0,
\end{equation}
with the normal vector~$\bfn$ oriented as in Fig.~\ref{fig:geometry}. For the
deformable air--liquid interface the boundary condition is
\begin{align}
  \label{EqBoundSur}
  \rho V^2 \partial_x^2 \phi  + \rho g \partial_z \phi - \gamma \partial_x^2 \partial_z \phi = 0
  \;\; (\text{at}\quad z = 0).
\end{align}
This boundary condition is obtained as follows. Using a height
profile~$\zeta(x)$, the kinematic boundary condition becomes for a nearly flat
interface ($\partial_x\zeta\ll1$),
\begin{equation}
  \label{kinemzeta}
  \partial_z\phi - \partial_x\zeta(V + \partial_x\phi)=0.
\end{equation}
Additionally, we have Laplace's law and Bernoulli's equation,
\begin{gather}
  \label{capillary}
  -p(x,\zeta(x)) + p^\air = \gamma\partial_x^2\zeta, \\
  \label{Bernoulli}
  \frac{\rho}{2}\bfv^2(x,\zeta(x)) + \rho g\zeta(x) + p(x,\zeta(x)) = \textit{const},
\end{gather}
with $\gamma$~the surface tension and with $p^\air$~the pressure above the
interface, which is assumed to be a constant. The curvature of the interface has
been linearized. The boundary condition~\eqref{EqBoundSur} is obtained by
inserting Laplace's law into the Bernoulli's equation, linearized in~$\phi$. A
derivative with respect to~$x$ and multiplying with~$\partial_x\phi$ allows to
eliminate the height profile~$\zeta$, using Eq.~\eqref{kinemzeta}. Terms which
are either quadratic in~$\phi$ or bilinear in $\phi$ and~$\zeta$ are neglected
as second-order perturbations.
\begin{figure}%
  \centering
  \includegraphics{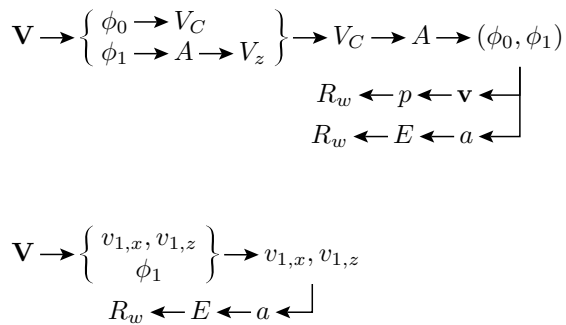}%
  \caption{Overview of the variable transformations of the full theory in
  Secs.~\ref{sec:flow} and \ref{sec:drag} (upper panel), and of the dipolar
  model in Sec.~\ref{sec:dipolar} (lower panel).}%
  \label{fig:organi}%
\end{figure}%

The problem we will solve in the following consists of the Laplace
equation~\eqref{incompress} for the unknown~$\phi(x,z)$, together with the two
boundary conditions~\eqref{kinemsphere} and~\eqref{EqBoundSur}. The flow
disturbance must further vanish at very large depths: $\partial_z \phi
\rightarrow 0$ when $z \rightarrow -\infty$. These equations can not be solved
directly by numerical means in real space, such as by a finite element method,
since the waves emitted by the cylinder propagate to infinity whereas any
discretization is done in a finite domain. Hence, additional analytic
transformations are needed to cast the problem in a more accessible form. An
overview of the following transformations is provided in Fig.~\ref{fig:organi}.
We split the potential into two components with different
physical origins $\phi = \phi_0 + \phi_1$. The term $\phi_0$~is dominated by the
perturbation stemming from the sphere, and $\phi_1$~is mainly the perturbation
from the free surface. Of course, both perturbations have to mutually respect
the presence of the other boundary. In more precise terms, the potential
$\phi_0$ obeys the Laplace equation with the boundary condition $\bfn\dotp
\bfnabla \phi_0 = -V_C(\theta)$ at the surface of the cylinder and $\nabla \phi_0
\rightarrow 0$ for $r \rightarrow \infty$.

The unknown function $V_C(\theta) = \mathbf{n}\dotp(\mathbf{V} + \nabla \phi_1)$
describes the flow created by $\mathbf{V}$ and the wake at the cylinder surface
(hence Eq.~\eqref{kinemsphere} is always verified). The flow~$\bfnabla\phi_1$
created by the surface-waves is oscillatory in the $x$~direction and decays
exponentially with the depth~$z$. These properties are naturally captured by the
integral representation commonly used in theories of surface
waves~\citep{Lighthill,RapGen96},
\begin{align}
  \label{phione}
  \phi_1(x, z) = \frac{1}{2 \pi} \int dk e^{i k x} e^{|k| z} A(k),
\end{align}
where $k$~denotes the wave number. The two unknown potentials $\phi_0(x, z)$ and
$\phi_1(x,z)$ are now replaced by the two unknown functions $V_C(\theta)$ and
$A(k)$. Those have to be determined self-consistently in order to satisfy the
boundary conditions Eqs.~\eqref{kinemsphere} and~\eqref{EqBoundSur}. This
procedure requires an expression of~$\phi_0$ in terms of~$V_C$, which is
given by the integral representation
\begin{align}
  \label{boundint}
  \phi_0(r, \theta) = -\frac{R}{\pi} \int \log | r e^{i \theta} - R e^{i \theta'} | V_C(\theta') d \theta' 
\end{align}
which uses the polar coordinates depicted in Fig.~\ref{fig:geometry}. Notice
that the origin of these polar coordinates coincides with the cylinder axis
while the origin of the Cartesian coordinates~$x,z$ is located at the free
interface above the obstacle. Relation~\eqref{boundint} is either obtained by
expanding~$V_C(\theta)$ as a Fourier series and by solving the Laplace equation
for each term. An alternative way is to use the full Green function of the
Neumann problem in the circle~\citep{Jackson75,Schindler06}. The logarithmic
integral kernel appears as a consequence the fundamental solution of the Laplace
equation in two dimensions. Notice that the part of the Green function which
enforces the Neumann boundary condition is also logarithmic.

The boundary condition~\eqref{EqBoundSur} leads to a first relation between
$V_C(\theta)$ and $A(k)$. Since it is invariant under translations in the
$x$~direction, it takes a simple form in Fourier space,
\begin{align}
  \label{eqA}
  A(k) &= -\frac{1}{|k|}\: \frac{\omega^2(k)+V^2 k^2}{\omega^2(k) - V^2 k^2} V_z(k)
  \quad\text{with} \\
  \label{defVz}
  V_z(k) &:= \int dx e^{-i k x} \partial_{z} \phi_0(x, 0)
\end{align}
The expression of $V_z(k)$ as a function of $V_C(\theta)$ is readily obtained by
inserting Eq.~\eqref{boundint} into the above Eq.~\eqref{defVz}. The integral
with respect to~$x$ can be evaluated analytically, which yields
\begin{equation}
  \label{Vztheta}
  V_z(k) = -R \int e^{-|k|(H-R\sin(\theta')} e^{-ikR\cos(\theta')}
  V_C(\theta')\:d\theta'
\end{equation}
The exponential factors arise from the Fourier transform of the derivatives of
the logarithmic kernel in Eq.~\eqref{boundint}. We have now expressed the
unknown function~$A(k)$ in terms of the other, $V_C(\theta)$. The remaining
kinematic boundary condition~\eqref{kinemsphere} at the cylinder surface now
serves as a closed equation for determining~$V_C(\theta)$,
\begin{align}
  &V_C(\theta) = \bfn\dotp(\bfV + \bfnabla\phi_1) \\
  &= \bfn\dotp\Bigl(\bfV - \frac{\bfnabla}{2\pi} \int dk\: \frac{e^{ikx}
  e^{|k|z}}{|k|}\: \frac{\omega^2(k){+}V^2 k^2}{\omega^2(k) {-} V^2 k^2} V_z(k)
     \Bigr)
\end{align}
By injecting $V_z$ from Eq.~\eqref{Vztheta} into the last integral and by
changing the order of integration, $V_C(\theta)$ is found to satisfy an integral
equation of Fredholm's second kind
\begin{align}
  \label{selfcons}
  V_C(\theta) = V \cos \theta + \frac{1}{2 \pi} \int K(\theta, \theta') V_C(\theta') d \theta',
\end{align}
with the kernel function
\begin{multline}
  \label{kernel}
  K(\theta, \theta') = R \int \bigl[i \; {\rm sgn}(k) \cos\theta + \sin \theta\bigr]
     \frac{\omega^2(k) + V^2 k^2}{\omega^2(k) - V^2 k^2}\\
     {}\times  e^{i k R (\cos \theta   - \cos \theta')}
      e^{-|k| (2 H - R \sin \theta - R \sin \theta')} \:dk
\end{multline}

In this integral, singular pole contributions appear when the denominator
vanishes, $\omega^2(k) - V^2 k^2=0$. This happens only for $V > c_\text{min}$.
The two positive solutions of this equation, which we call~$k_g$ and $k_c$
($k_c>k_g$) correspond to gravity waves and capillary waves, respectively. In
the far-field regimes, they dominate the flow and thus correspond to the
inverse wavelengths found before and behind the obstacle. Directly above the
cylinder no unique wavelength can be identified. In order to ensure that waves
only leave the object and are not coming back from infinity, we introduce an
infinitesimal imaginary part into the dispersion relation. This term can be
understood as an infinitesimal viscous term~\citep{RapGen96,CheRap03}.
With the correct choice of its sign, the denominator becomes
$\omega^2(k)-V^2k^2-i\varepsilon k$. Notice that this choice also ensures that
capillary waves are emitted to the front and that gravity waves rest at the
rear of the obstacle.

\section{Wave drag}
\label{sec:drag}

Provided that the self-consistent Eq.~\eqref{selfcons} has been solved to find the
function~$V_C$, one can reconstruct the flow everywhere in the domain around
the obstacle. This procedure requires to calculate in turn $\phi_0$, $A(k)$,
and $\phi_1$, using Eqs.~\eqref{boundint}, \eqref{eqA}, and \eqref{phione}.
Once the flow is known, two different strategies offer themselves for the
calculation of the wave drag. Both are indicated in the upper part of
Fig.~\ref{fig:organi}. The first, which is conceptually simpler,
passes over the pressure field~$p(x,z)$, using Bernoulli's equation
\begin{equation}
  \frac{\rho\bfv^2(x,z)}{2} + \rho g z + p(x,z) = \frac{\rho V^2}{2}.
\end{equation}
The pressure field can then be integrated around the surface of the cylinder to
find the total drag force. In an inviscid fluid, d'Alembert's theorem ensures
that the drag caused by the emitted surface waves is the only contribution to the
drag force~\citep[\textsection~11]{LanLif59}. The wave drag then reads: 
\begin{equation}
  R_w = -\frac{1}{V} \mathbf{V}\dotp \oint_{S_R} dA(x,z)\, p(x,z)\, \mathbf{n}.
\label{rwpres}
\end{equation}
The second approach to calculate the wave drag makes use of the power carried
away by the waves. For the linear waves used here, the relation between this
power and the amplitude of the velocity oscillations in the far-field regime are
well known~\citep{RapGen96}. The oscillations in the far-field regime are of the
form $\zeta(x) = a\cos(kx + h)$. The amplitudes~$a_c, a_g$, the wave numbers
$k_c, k_g$ and the phases $h_c, h_g$ are different for capillary and for gravity waves.
Each energy density of the particular wave depends quadratically on the
amplitude,
\begin{equation}
  \label{Ea}
  E(k) = \frac{1}{2}\rho c(k)^2\,k a^2.
\end{equation}
This energy density may alternatively be expressed in terms of the amplitude of
the velocities, denoted by $v_{z,\infty}$, which is related to the amplitude $a$
of the deformation by $v_{z,\infty} = V k a$. This relation is a consequence of
the linearized kinematic boundary condition~\eqref{kinemzeta}. It allows to
determine the amplitude~$a$ from the flow, once the function~$V_C(\theta)$ is
known. The two energy densities allow to calculate the power carried away by the
waves, from which the drag results as~\citep{RapGen96},
\begin{equation}
  R_w = \frac{V-\omega'(k_c)}{V} E(k_c) - \frac{V-\omega'(k_g)}{V} E(k_g).
  \label{rwenergy}
\end{equation}
The energy of each wave is transported at the group velocity~$\omega'$.

The two expressions \eqref{rwpres} and \eqref{rwenergy} for the wave drag are
equivalent. Their comparison allows to verify the accuracy of the numerical
solutions below.

\section{Numerical results}%
\label{sec:num}%
\begin{figure}%
  \centering
  \includegraphics[width=\linewidth]{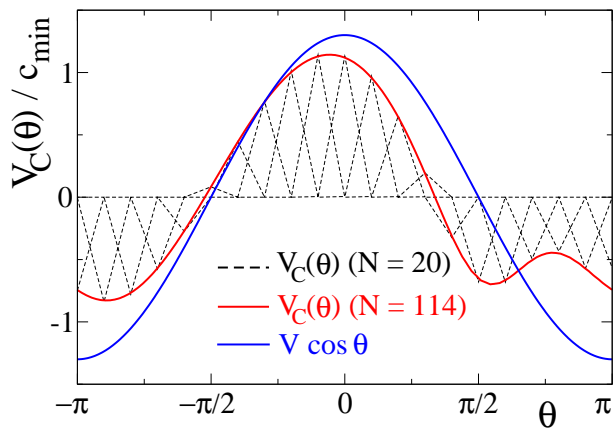}%
  \caption{(Color online) Numerical solution of Eq.~\eqref{selfcons}. The
  dashed lines represent the discretisation with linear finite elements.
  Parameters: $H\kappa=0.5$, $R\kappa=0.3$, and $V = 1.3c_\text{min}$.}%
  \label{fig:fem}%
\end{figure}%
The self consistent equation Eq.~(\ref{selfcons}) can be solved by
numerical methods. Our first step is to create a numerical table of the values
of the kernel~$K(\theta, \theta')$, the integral in the definition of the kernel
is computed using standard numerical routines from the GSL~library~\citep{gsl}.
It provides special discretizations for calculating principal values which
appear in Eq.~\eqref{kernel} due to the singular integrand. The equation
Eq.~(\ref{selfcons}) is then discretized using finite elements. The
function $V_C(\theta)$ is approximated using a finite set of $N$~basis functions
$\phi_n(\theta)$.
\begin{align}
  V_C(\theta) = \sum_{n=1}^{N} c_n \phi_n(\theta)
  \label{eq:VcApprox}
\end{align}
For simplicity we have chosen a basis of $N$~piecewise linear hat functions with
periodic boundary conditions on the $(-\pi, \pi)$ interval. Fig.~\ref{fig:fem}
shows an example of such an approximation for a small number of elements
$N=20$. This expression for~$V_C(\theta)$ is then inserted in
Eq.~(\ref{selfcons}) and Galerkin's method is used to convert it into a matrix
equation of size $N \times N$. Namely we multiply  Eq.~(\ref{selfcons}) by one
of the function~$\phi_m(\theta)$ and integrate over the angle~$\theta$. This
leads to a linear equation on the coefficients~$c_n$:
\begin{align}
  \sum_n [ (\phi_m, \phi_n) - \frac{1}{2 \pi} (\phi_m, K \phi_n) ]c_n = V (\phi_m, \cos \theta) 
  \label{eqfem}
\end{align}
where we have introduced the notation 
$K \phi_n(\theta) = \int K(\theta, \theta') \phi_n(\theta') d\theta'$ and 
defined scalar product between two arbitrary functions $f(\theta)$ and $g(\theta)$ as
\begin{align}
  (f,g) := \int f(\theta) g(\theta) d\theta 
\end{align}
In a last step the system Eq.~(\ref{eqfem}) is solved using a LU decomposition 
\citep{numres}. The integrals involved in the calculation of the scalar products
are determined numerically~\citep{gsl}. It is possible to check the convergence
of our numerical procedure by inserting the obtained approximation
for~$V_C(\theta)$ into Eq.~(\ref{selfcons}). The degree of accuracy can then be
estimated from the difference between the left and right hand side of
Eq.~(\ref{selfcons}). For the typical number of elements we use in our
simulations $N \simeq 110$ the relative difference is of the order of~$10^{-3}$. 

Figure~\ref{fig:fem} presents the solution of Eq.~(\ref{selfcons}) for the
geometrical parameters $H\kappa=0.5$, $R\kappa=0.3$ and a flow
velocity of $V = 1.3 c_\text{min}$. It is compared with the source term of the
integral equation $V \cos \theta$ which results
if the contribution from the surface waves is neglected. As can be
seen, the function~$V_C(\theta)$ is strongly modified by the
self-consistent interaction between the cylinder and the emitted
capillary-gravity waves.%
\begin{figure}%
  \centering
  \includegraphics[width=\linewidth]{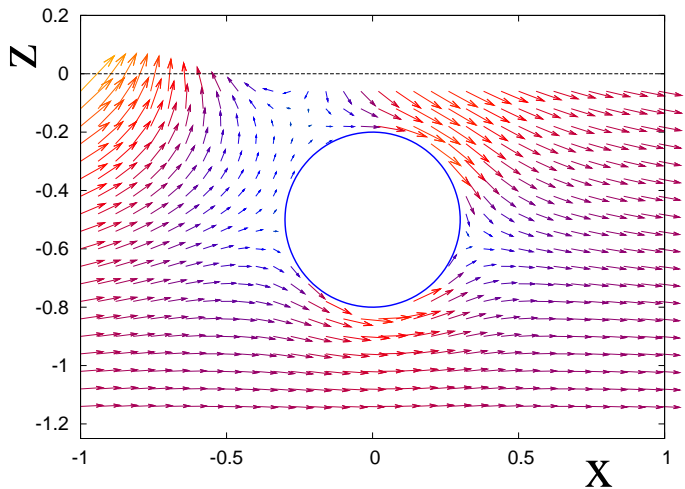}%
  \caption{(Color online) Visualization of the flow velocity around the
  cylinder. The velocities are represented by arrows the size of which is
  proportional to the velocity magnitude. The color code is yellow/gray for
  large magnitudes and blue/black of small ones. Parameters as in
  Fig.~\ref{fig:fem}.}%
  \label{fig:flow}%
\end{figure}%
\begin{figure}%
  \centering
  \includegraphics[width=\linewidth]{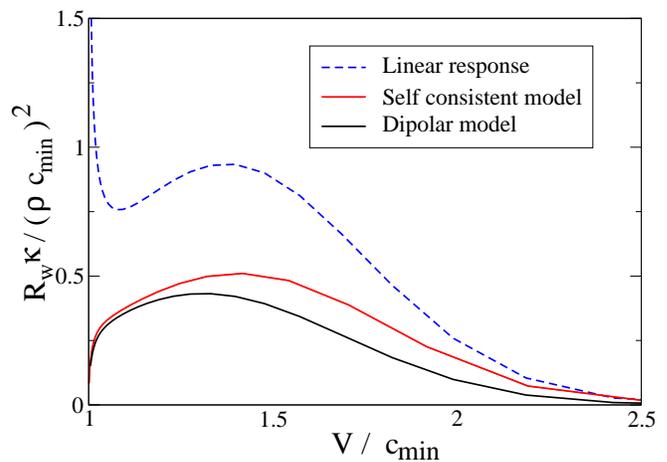}%
  \caption{(Color online) Wave drag as a function of the far-field velocity~$V$
  for the parameters for $ H \kappa = 0.5$ and $ R \kappa = 0.3$. The results of
  three different models are presented: the dashed curve corresponds to a linear
  response calculation where $V_C = V \cos \theta$, the red (gray) curve
  corresponds to a self-consistent calculation using Eq.~(\ref{selfcons}) and
  the black curve corresponds to the dipolar model explained in section IV.}%
  \label{fig:drag}%
\end{figure}%

Once the function $V_C(\theta)$ has been determined we can calculate the flow in
all the space $z < 0$ around the cylinder following the steps described in
section II. The velocity field around the cylinder for the parameters of
Fig.~\ref{fig:fem} is depicted in Fig.~\ref{fig:flow}. Notice that the flow
obeys the kinetic boundary condition on the sphere given by
Eq.~(\ref{kinemsphere}) which confirms our numerical procedure. 

Using the results from Sec.~\ref{sec:drag}, we can now calculate the wave drag
as a function of the externally applied flow velocity~$V$. Figure~\ref{fig:drag}
shows this function, as determined numerically from both, the direct integration
of the pressure field around the sphere~\eqref{rwpres}, and from the energy
balance in Eq.~\eqref{rwenergy}. Their difference in Fig.~\ref{fig:drag} is
smaller than the line width. At velocities~$V$ close to the critical
value~$c_\text{min}$, the drag vanishes continuously. This behavior is
qualitatively different from that found from linear response theory with an
external pressure field as perturbation, as it predicts a divergence
at~$c_\text{min}$ ~\citep{RapGen96}. Our approach reduces to this linear
response theory if, instead of solving the self consistent Eq.~(\ref{selfcons}),
$V_C(\theta)$ is set to the result $V \cos \theta$ for a cylinder moving in a
quiescent inviscid liquid without surface. The wave drag obtained in this case
diverges close to $V = c_\text{min}$ as shown by the dashed curve on
Fig.~\ref{fig:drag}. At large velocities, the wave drag determined from the self
consistent model  curve rejoins the curve from linear response.

The observation in Fig.~\ref{fig:drag}, that the drag is a continuous function
at the critical velocity, constitutes the central result of the present work.
The regularization is achieved by the self-consistent treatment of the emitted
waves, which takes into account the mutual interaction between the perturbations
created by the cylinder and the waves, respectively. The physical origin of the
regularization can be understood with a simpler model which does not require the
full solution of the self-consistent equation. This model will be detailed in
the following section, where we derive a square-root scaling for the wave drag
close to~$c_\text{min}$.

\section{Dipolar model}
\label{sec:dipolar}

We now treat the interaction between the wake and the cylinder in an approximate
manner. Instead of enforcing the exact kinematic boundary
condition~\eqref{kinemsphere} at all points on the cylinder, we impose it only
as an averaged constraint. This approach assumes that the flow~$\bfv_1 =
\bfnabla\phi_1$, created by the waves, is homogeneous around the cylinder. The
response of the cylinder is now reduced to a simple dipolar potential which
describes the response to a yet unknown uniform flow~$(\bfV+\bfv_1)$,
\begin{equation}
  \label{}
  \phi_0 = R^2 (\bfV + \bfv_1)\dotp\bfnabla\log r.
\end{equation}
The name ``dipolar'' arises from the analogy with the potential of a dipole in
two-dimensional electrostatics. This approximation reduces the complexity of
the equation~\eqref{boundint}. Figure~\ref{fig:organi} summarizes the necessary
transformations in the same way as it was done for the sections above. Instead
of determining a whole function~$\phi_0$ we must now only find two
parameters~$v_{1,x}$ and~$v_{1,z}$. With these parameters, the function
$V_C(\theta)$ from Sec.~\ref{sec:flow} assumes the form $V_C(\theta) = (V +
v_{1,x}) \cos \theta + v_{1,z} \sin\theta$, which can be used immediately to
calculate the unknown function~$A(k)$. The self-consistent
equation~\eqref{selfcons} for~$V_C$ is thus replaced by a linear set of two
equation for the parameters. The solution is
\begin{equation}
  \left(\begin{array}{c}
  V + v_{1,x} \\
  v_{1,z}
  \end{array} \right) =  \frac{V}{(1-I_1)^2 + I_2^2 }  \left( \begin{array}{c}
  1 - I_1\\ - I_2
  \end{array} \right),
\end{equation}
with the shortcuts
\begin{align}
  I_1 &= \int \frac{d k}{2 \pi} \pi R^2 e^{-2 H |k|} |k| \frac{\omega^2(k) + V^2 k^2}{\omega^2(k) - V^2 k^2}, \\
  I_2 &=  \int \frac{d k}{2 \pi} \pi R^2 e^{-2 H |k|} i k \frac{\omega^2(k) + V^2 k^2}{\omega^2(k) - V^2 k^2}.
\end{align}
In order to obtain the wave drag~$R_w$ from this result, we follow the same
steps as in Sec.~\ref{sec:drag}. This time, we can do them
analytically, not only numerically, thanks to the simple form of the
potential~$\phi_0$. In this treatment, we prefer the energy balance argument
instead of the pressure integral. In the far-field, $x\to\pm\infty$, we find the
wave amplitude as follows,
\begin{equation}
  \label{dipol:a}
  a = 2 \pi R^2 \frac{k V e^{-H |k|}}{|\omega'(k)-V|}
  \frac{1}{\sqrt{(1-I_1)^2 + I_2^2}}\;,\;c(k)=V
\end{equation}
In the course of this calculation, we pass from $\phi_0$ to $A(k)$, using
Eqs.~\eqref{eqA} and~\eqref{defVz}. In the far field regime the only
contribution to the flow arises from the waves. The flow potential $\phi_1$
created by the waves is given by a Fourier-Laplace transform of $A(k)$ (see
Eq.~\eqref{phione}). In the asymptotic regime $|x| \rightarrow \infty$ we can keep only the
contributions from the emitted waves of fixed wave numbers $k = k_c, k_g$. They
appear from the delta function contribution of the integration around the poles
of $A(k)$, using the identity~\citep[p.~481]{MatWal70}:
\begin{align}
  &\lim_{\varepsilon\to0} \frac{1}{\omega^2(k) - V^2k^2 - i\varepsilon k} \nonumber \\
  &= \textrm{P.v.}\frac{1}{\omega^2(k) - V^2k^2} + i\pi\textrm{sgn}(k)\,\delta\bigl(\omega^2(k)
  - V^2k^2\bigr),
  \label{identity}
\end{align}
where $\textrm{P.v.}$ denotes the Cauchy principal value and the delta function can be 
transformed as:
\begin{equation}
  \delta\bigl(\omega^2(k) - V^2k^2\bigr)
  = \sum_{k_i:c(k_i)=V} \frac{\delta(k{-}k_i)}{2 V |k_i|\: |\omega'(k_i)-V|}.
\end{equation}

Now, that we have an expression for the deformation amplitude~$a$, we can find
the wave drag using Eqs.~\eqref{Ea} and \eqref{rwenergy}. The auxiliary
integrals $I_1$ and $I_2$ are calculated numerically. The result are displayed
in Fig.~\ref{fig:drag} and show that the dipolar model reproduces the behavior
of the exact solution from Sec.~\ref{sec:flow}. The accuracy of the dipolar
model can be understood because the radius of the cylinder is small compared to
the capillary wave length $R = 0.3 \kappa^{-1}$. Hence the flow is reasonably
uniform on this scale. A good agreement is found when $V \to c_\text{min}$
because the emitted wavelengths tend both to the same value $\kappa^{-1}$, which
is larger than the object size. At higher velocities, where the emitted
wavelengths differ, the agreement decreases because the shorter one may become
smaller than the object size.

Our aim now is to find the square-root behavior of the drag force,
$R_w\propto\sqrt{V-c_\text{min}}$ in the frame of the dipolar model. In order to
do so, we  analyze the scaling of the wave amplitude in Eq.~\eqref{dipol:a} in
the vicinity of $c_\text{min}$. We then use Eqs.~\eqref{Ea} and \eqref{rwenergy}
to establish the scaling for the wave drag. All scaling behavior near
$c_\text{min}$ stems from the scaling of $|\omega'(k)-V|$ where $k$ obeys $c(k)
= V$. Near $k=\kappa$ the phase velocity shows a quadratic minimum and the
selected wave numbers are: 
\begin{equation}
  k_c, k_g \simeq \kappa \left(1 \pm 2 \sqrt{
    \frac{V-c_\text{min}}{c_\text{min}}} \right).
\end{equation}
The difference $|\omega'(k)-V| = |\omega'(k)-c(k)|$ is thus determined by
the linear contribution from the group velocity which is given by: 
\begin{equation}
  |\omega'(k)-V| \simeq \sqrt{c_\text{min}(V - c_\text{min})} 
\end{equation}
Note that the result is the same for both capillary and gravity waves. 

The scaling of $|\omega'(k)-V|$ is taken over to the amplitude $a$, to $I_2$ and
to the wave drag $R_w$. The term $I_1$ becomes constant in the limit $V\to
c_\text{min}$, while the term~$I_2$ diverges. Indeed $I_1$ involves only the
principal value term from Eq.~\eqref{identity} while $I_2$ depends on the delta
function part that scales as $1/|\omega'(k)-V|$. After all these preparations, we
are now ready to provide the scaling of the wave drag:
\begin{equation}
  R_w \simeq \rho c_\text{min}^2 \kappa^{-1} \sqrt{ \frac{V-c_\text{min}}{c_\text{min}} }
\end{equation}
This expression is derived assuming $R \ll H \ll \kappa^{-1}$, and it holds for
velocities very close to $c_\text{min}$, where $I_2 \gg 1$ (this occurs as soon
as $V - c_\text{min} \ll ( \kappa^2 R^2 )^2 $).

While our approach is exact for a linear capillary-gravity waves non linear
corrections become relevant when the flat interface condition $\partial_x \zeta
\ll 1$ is not verified anymore~\citep{Dias}. We have checked numerically the validity of this
assumption for the parameters of Fig.~\ref{fig:drag}. It is clearly verified at
large velocities but as $V$ approaches $c_\text{min}$ the typical $\partial_x \zeta$
increases and reaches a regime where $\partial_x \zeta \simeq 1$ for $V \simeq
c_\text{min}$. In this range, where we find moderate nonlinearities of the order one,
we expect that they will change only the quantitative predictions of the
self-consistent theory, without introducing qualitatively new features. This is
consistent with the observation that nonlinear waves (for e.g.~solitons) are not
seen in the experiments close to $c_\text{min}$ and that form of the generated wave
pattern is well described in linearized theory~\cite{BurSte02}. Hence although
nonlinearities should be taken into account for a complete hydrodynamic theory,
the regularization of the wave drag near $V \simeq c_\text{min}$ is already present
in the linear self-consistent theory. Notice that approaches that include only
nonlinear effects without a self-consistent treatment fail to produce stable
solutions near~$V=c_\text{min}$~\citep{ParVanCoo05}.

Recently, \citet{Rabaud} developed an experimental technique that allows to
recover the height profile created by a moving disturbance. This technique is a
promising candidate for the verification of our theoretical predictions on the
onset of the wave drag and the implied wave pattern close to~$c_\text{min}$.

\section{Conclusion}

We have addressed the behavior of the wave drag for objects of extension smaller
and close to the capillary length moving at a speed close to the critical
velocity $c_\text{min}$, which is given by the minimum of the wave velocity for
capillary gravity waves. It is known that theories where the object is modeled
by an external pressure source lead to singular wave drag behavior at $V =
c_\text{min}$. In this article, we show that even for linear capillary-gravity waves,
this singularity can be removed, even in the approximation of an inviscid fluid,
by treating the boundary conditions at the object interface in an exact way. For
this purpose, we treat the wave emission problem by a cylinder submerged near
the liquid-air interface under an external flow. We derive a self-consistent
integral equation describing the flow velocity at the cylinder interface. This
equation is solved numerically with a finite elements method, which allows us to
reconstruct the flow in the entire space around the cylinder and to determine
the wave drag on the cylinder. In addition to the numerical solution, we propose
a simple approximation valid for cylinder diameters smaller than the wavelength,
where the interaction between the waves and the cylinder is treated in a dipolar
approximation. In this case it is possible to make analytic estimates showing
that $R_w \propto \sqrt{V - c_\text{min}}$.

Our findings explain why a smooth onset of the wave drag is observed even if the
shape of the wave pattern is well described by linear response. The validity of
our theory is limited by the validity of the flat interface approximation. In
our simulations, this conditions is not well verified for velocities
around~$c_\text{min}$. In principle, nonlinear corrections should therefore become
relevant. However, we think that the qualitative behavior is already captured by
our self-consistent theory, even if the inclusion of nonlinear effects would be
required for quantitative predictions.

One might be tempted to compare the scaling $R_w \propto \sqrt{V -
c_\text{min}}$, which we found in the dipolar model, with the square-root fit
done by \citet{BurSte02} in their Fig.~18. This apparent agreement between
theory and experiment must be taken with care. Our results are valid for a long
cylindrical obstacle, while the experiment was done with a spherical object.
Moreover, the experimental data does not allow to determine unambiguously a
square-root scaling at the onset of the wave drag. Hence, more experiments and a
three-dimensional theory will be required to establish the exact scaling at the
transition. The main point however, namely the fact that our theory is able to
recover the continuous onset of the wave drag at $c_\text{min}$, coincides with
the mentioned experiment. Our finding of a continuous drag force might serve as
an element of understanding the motion of small insects on or near water
surfaces, since such animals have to find a delicate balance between viscous and
wave drag.

\begin{acknowledgments}
We would like to thank F.~Closa and M.~Rabaud for interesting discussions.
A.~D.~Chepelianskii acknowledges~DGA for support.
\end{acknowledgments}

\end{document}